\documentclass[
aps,prd,
showpacs,twocolumn,notitlepage,
amssymb,amsmath,amsfonts,
nofootinbib,superscriptaddress,
floats,
amsmath,amssymb,
aps,
]{revtex4-2}

\usepackage{aas_macros}
\usepackage{graphicx}% Include figure files
\usepackage{dcolumn}% Align table columns on decimal point
\usepackage{bm}% bold math
\usepackage{xcolor}
\newcommand{\red}[1]{\textcolor{black}{#1}}

\newcommand{\blue}[1]{\textcolor{black}{#1}}

\begin{document}

%\preprint{APS/123-QED}

\title{
Connecting GRBs from Binary Neutron Star Mergers to
Nuclear Properties of Neutron Stars
} 

\author{Rosalba Perna}
\email{rosalba.perna@stonybrook.edu.}
 \affiliation{Department of Physics and Astronomy, Stony Brook University, Stony Brook, NY 11794-3800, USA} 
\author{Ore Gottlieb}
\affiliation{Center for Computational Astrophysics, Flatiron Institute, New York, NY 10010, USA}
    \affiliation{Department of Physics and Columbia Astrophysics Laboratory, Columbia University, Pupin Hall, New York, NY 10027, USA}
    
\author{Estuti Shukla}
\affiliation{Institute for Gravitation and the Cosmos, The Pennsylvania State University, University Park PA 16802, USA}
\affiliation{Department of Physics, The Pennsylvania State University, University Park PA 16802, USA}

\author{David Radice}
\affiliation{Institute for Gravitation and the Cosmos, The Pennsylvania State University, University Park PA 16802, USA}
\affiliation{Department of Physics, The Pennsylvania State University, University Park PA 16802, USA}
\affiliation{Department of Astronomy \& Astrophysics, The Pennsylvania State University, University Park PA 16802, USA}

\date{\today}% It is always \today, today,
             %  but any date may be explicitly specified

\begin{abstract}
The fate of the binary neutron star (NS) merger remnants hinges sensitively upon the NS equation of state and the threshold mass, $M_{\rm ls}$, that separates a long-lived from a short-lived NS remnant.
The nature of the electromagnetic counterparts is also influenced by the remnant type, particularly in determining whether a gamma-ray burst from a compact binary merger (cbGRB) is of short or long duration. We propose a novel approach to probe  \blue{$M_{\rm ls}$}
by linking it to the estimated observed ratio of long to short cbGRBs. We find that current observations broadly favour a relatively high value for this transition, $M_{\rm ls}\simeq 1.3 M_{\rm TOV}$, for which $ M_{\rm TOV} \lesssim 2.6\,M_\odot $, consistent with numerical simulations, as also shown here.
Our results disfavour nuclear physics scenarios that would lead to catastrophic pressure loss at a few times nuclear density and temperatures of tens of MeV, leading to a rapid gravitational collapse of binaries with total mass $M \lesssim 1.3 M_{\rm TOV}$.
Future individual gravitational wave events with on-axis cbGRBs can further bound $M_{\rm ls}$.

\end{abstract}

                              
\maketitle

\section{\label{sec:intro}introduction}

The joint gravitational wave (GW) and electromagnetic (EM) detections of the binary neutron star (BNS) merger of GW170817 \citep{Abbott2017} confirmed the long-standing hypothesis that BNS mergers are the progenitors of short gamma-ray bursts (GRBs) \citep{Mooley2018}. GW170817 highlighted the unique capacity of observations to constrain the neutron star (NS) equation of state (EoS), based on the properties of the compact object remnant following the merger \citep{Abbott2017,Abbott2018,Margalit2017,Shibata2017,Most2018,Radice2018,Radice2018b,Rezzolla2018,Shibata2019,Dietrich2020}. Additional constraints on the EoS may also emerge from the jet characteristics influenced by interactions with merger ejecta, whose properties depend on the EoS \citep{Lazzati2019}.

Short GRBs, with an average duration of $\lesssim 1$~seconds, have traditionally been regarded as a distinct class from long GRBs, which typically exhibit an average duration of $\sim 30$~seconds \citep{Kouveliotou1993} and are generally attributed to collapsar events \citep{Woosley1993,MacFadyen1999}. The separation between these classes is conventionally set at 2 seconds. However, recent optical and infrared detections of kilonovae in two $\sim 10$-second-long GRBs \citep{Rastinejad2022,Troja2022,Yang2022,Yang2023,Zhang2022,Levan2023b,Sun2023} suggest that binary mergers can produce durations well within the range traditionally associated with long GRBs, exceeding the typical accretion timescale of a disk previously thought to set the short GRB duration. Although specific cases may be attributed to unusual mechanisms (e.g., the 64-second GRB191019A at low redshift, which lacked an accompanying supernova but was shown to be consistent with an intrinsically short GRB from a high-density region, \cite{Lazzati2023}), the growing body of observations calls for a more fundamental explanation of this new GRB class.

\citet{Gottlieb2023} proposed a unification model for short- and long-duration GRBs resulting from binary mergers (sbGRBs and lbGRBs, respectively). This model bases the distinction on the nature of the compact remnant and the disk mass. In this model, the new class of lbGRBs emerges from black holes (BHs) with massive ($M_d\gtrsim 0.1 M_\odot$) post-merger accretion disks. These may form either following a short-lived  NS remnant or a prompt-collapse BH in a moderately unequal mass BNS or BH--NS mergers. By contrast, sbGRBs result from long-lived remnant NS or BHs with less massive disks. Although jets powered by NSs and BHs may appear similar, \citet{Gottlieb2024} proposed that the kilonovae associated with sbGRBs can help distinguish the central engine.

A recent analysis by \citet{Rastinejad2024} indicates that lbGRBs are accompanied by bright red kilonovae, consistent with an origin in massive BH disks, while sbGRBs are linked to luminous blue kilonovae. \citet{Gottlieb2024} showed that the brightness and color of kilonovae associated with sbGRBs cannot be accounted for by BH-powered jets alone. They concluded that long-lived remnant NSs are likely the central engines of sbGRBs, whereas low\blue{-mass} BH disks rather power fainter lbGRBs. In fact, they pointed out that GW170817 was likely an lbGRB rather than an sbGRB.

The framework of \citet{Gottlieb2024}, which connects diverse GRB classes from BNS mergers with kilonova properties, remnant types, and post-merger accretion disk masses, also enables the extraction of essential information about the NS EoS. 
\blue{The connection between remnant types from NS-NS mergers, short GRBs, and the NS EoS has long been noted in the literature. \citet{Belczynski2008} and \citet{Fryer2015} pointed out that $M_{\rm TOV}$ can be constrained under the assumptions that short GRBs from NS-NS mergers require a BH remnant. \citet{Salafia2022} discussed a method to constrain the NS EoS and mass distribution using the fraction of GW NS-NS mergers with an associated jet, with the latter assumed to require a BH remnant and a minimum disk mass of $\sim 10^{-3} M_\odot$. } 
\blue{The statistical} approach \blue{in terms of incidence of remnant types}
has \blue{also} been previously utilized by \citet{Piro2017,Sarin2020,Guglielmi2024}, who constrained the EoS based on the fraction of short GRBs with plateaus, under the assumption that these plateaus indicate a remnant NS passing through a magnetar phase either as a stable NS, or a NS remnant collapsing to a BH after losing its centrifugal support.

Here, building on the recent unification model by \citet{Gottlieb2023,Gottlieb2024} and current observational constraints on the relative rates of lbGRBs to sbGRBs, we provide a novel approach to constraining nuclear properties of NS matter which has particular sensitivity to the threshold NS  mass setting the transition between a long- and short-lived NS remnant. The paper is organized as follows. In Section~\ref{sec:methods},
we describe the key model ingredients of our analysis: the intrinsic distribution of NS masses, the binary NS merger remnant types, and the modeling of the disk mass remnant as a function of the total BNS mass and their mass ratio. 
In Section~IIIA, we present the results of our population study, focusing on relative fractions of the merger remnants and their surrounding disks. These are transformed in Section~IIIB  into theoretical predictions on the relative rates of lbGRBs to sbGRBs according to the model of \citet{Gottlieb2023,Gottlieb2024}. We compare these predictions with observations, to constrain the threshold mass between a long-lived and a short-lived NS remnant. In Section~IIIC, we compare our results to numerical simulations and find a remarkable consistency.
We conclude and highlight the implications of our results for the nuclear properties of NS matter in Section IV.

\section{\label{sec:methods}Key Model Ingredients}

\subsection{Intrinsic distributions of NS masses}

The mass distribution of the BNS merger remnants closely hinges upon the mass distribution of the NS binary components. 
Galactic NS star systems are known to have a distribution tightly peaked around $1.35 M_\odot$, but the mass distribution of all known pulsars is now known to be broader \cite{Tauris2017}, with the current range spanning from $1.17 M_\odot$ to $2.35\pm 0.17 M_\odot$ \cite{Romani2022}.
As the wealth of observations has been growing over the years combining  measurements from different types of systems, it has become 
apparent that the NS mass distribution is bimodal \cite{Schwab2010,Antoniadis2016,Alsing2018,Farr2020,Galaudage2021,Rocha2023}, consistent with what was found in simulations of the explosions of massive stars \cite{Zhang2008}. 

\begin{figure}
    \centering
    \includegraphics[width=1.05\linewidth]{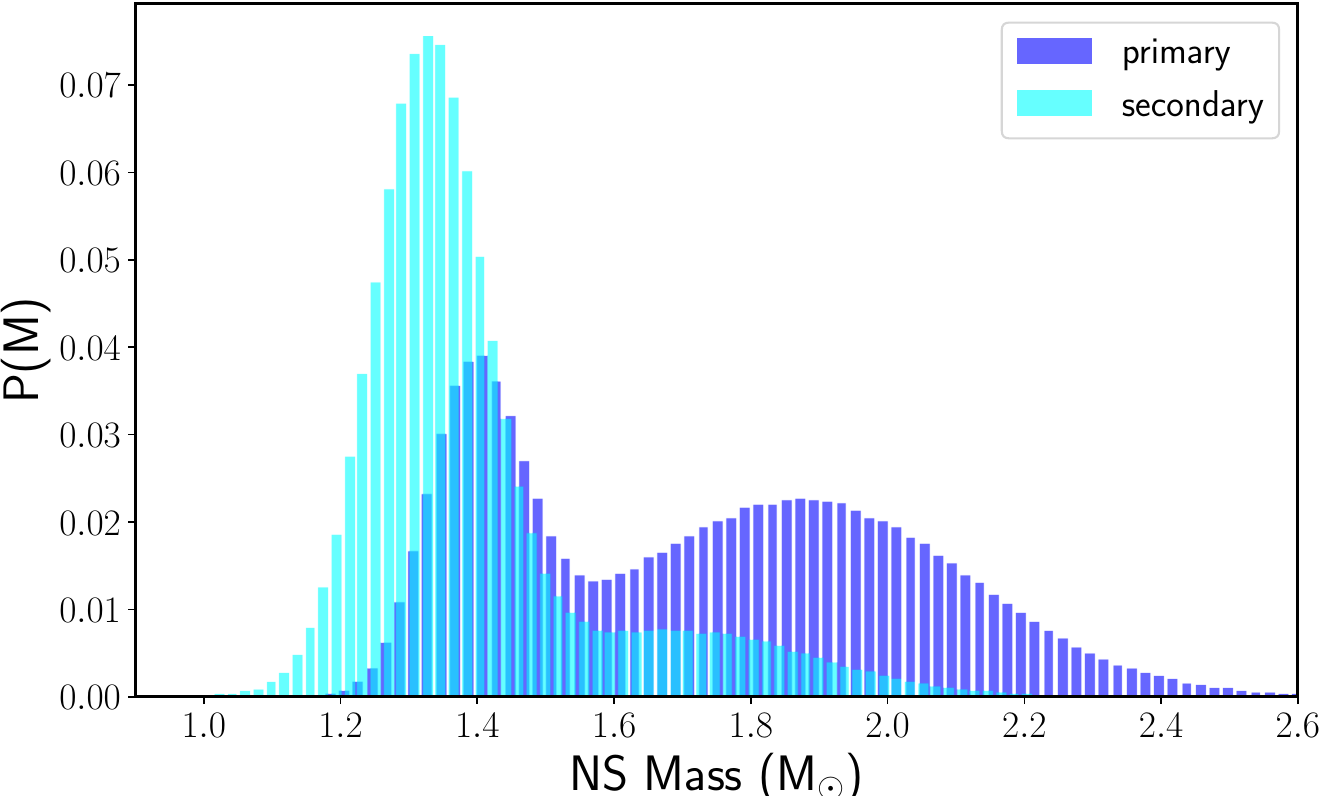}\\
    \includegraphics[width=1.05\linewidth]{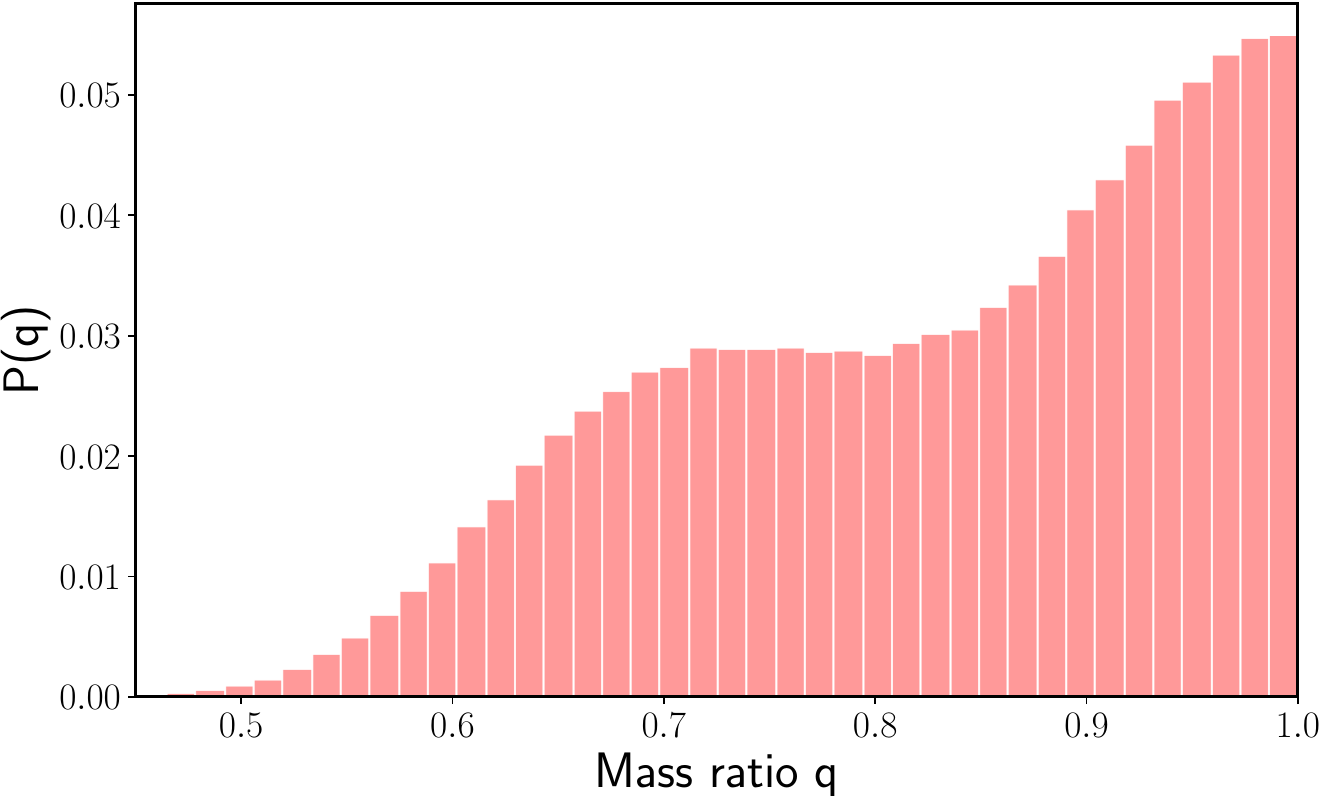}
    \caption{{\em Top:} Probability distributions of primary ($M_1$) and secondary ($M_2$) NS mass.
    {\em Bottom:} Corresponding probability distribution of the mass ratio $q$ under a \red{random pairing} assumption on their mass ratio.}
\label{fig:masses}
\end{figure}

Here we adopt the best-fit parameters from the recent study by Rocha {\it et al.} 
\cite{Rocha2023}, who analyzed the most up-to-date sample of binary systems in the Galaxy and in Globular Clusters.
They fitted the sample with a double Gaussian with best-fit
parameters: $\mu_1=1.351 M_\odot$, $\sigma_1=0.084 M_\odot$,
$A_1=0.539$ for the mean, standard deviation, and normalization of the
lower-mass peak, respectively, and $\mu_2=1.816 M_\odot$, $\sigma_2=0.260 M_\odot$,
$A_2=0.460$ for the second, higher-mass peak.  

Lacking detailed predictions for the relative mass distributions of the two NS members of the binary (i.e. their mass ratio), and to avoid making assumptions that then influence the results,  we choose to make no prior assumptions on the mass ratio:
both NS masses are randomly generated from the same binary
distribution \footnote{\red{A note of caution is that, for main sequence stars in binaries, observations show that random pairing is not supported \cite{Kouwenhoven2009}. For NS in binaries, more data will be needed to draw definite conclusions.}}, with the primary being the larger of any two randomly drawn values. The resulting distributions of the primary ($M_1$) and secondary ($M_2$) 
masses are displayed in the top panel of 
 Fig.~\ref{fig:masses}, while their mass ratio $ q \equiv M_2/M_1 $ is correspondingly shown in the bottom panel of the same figure.
It can be seen that, 
 with the \red{random pairing} assumption made on the $q$ distribution, the natural probability favors comparable masses.

\blue{This is the NS mass model we designate as our fiducial choice. In the appendix, we will explore how varying this model in different ways affects our results. }

\subsection{\blue{Merger outcome} as a function of the EOS}

Given two merging NSs of masses $M_1$ and $M_2$ (randomly drawn from
the distributions above), we calculate the total  mass \blue{of the binary when the stars are at infinite separation} as $M_{\rm
  tot}=M_1+M_2$. \blue{ We note that this should not be intended as the baryonic or gravitational merger remnant often used in the literature (e.g. \cite{Piro2017,Shibata2019,Salafia2022}).}

There are five possible \blue{outcomes} from a BNS merger \cite{Radice2020}:

\begin{itemize}
    \item[(a)] {\em Stable Neutron Star} (SNS): If $M_{\rm tot}\le M_{\rm TOV}$, where $M_{\rm
  TOV}$ is the maximum mass of a cold, non-rotating NS. 
  
  \item[(b)] {\em Very Long-lived Neutron Star} (\blue{VLNS}): If $ M_{\rm TOV} < M_{\rm tot} \le M_{\rm L,sp}$,
where $M_{\rm L,sp}$ is the maximum mass of a uniformly rotating
NS secured by centrifugal and thermal effects that possibly
collapses to a BH on a dynamical or secular timescale. 
\blue{While it has been shown that rigid rotation can support stars with mass
up to 1.2 $M_{\rm TOV}$ \cite{Breu2016}, here}
we use the lower-limit $M_{\rm L,sp}= 1.15M_{\rm
  TOV}$. \blue{This is a conservative limit for the results presented in Sec.~III, as quantitatively shown in the appendix.} 

\item[(c)] {\em Long-Lived Neutron Star} (\blue{LLNS)}:
If $ M_{\rm L,sp}\lesssim M_{\rm tot}< a M_{\rm TOV}$. These are long-lived \blue{objects} that collapse on the viscous timescale.
The mass $M_{\rm ls}\equiv a M_{\rm TOV}$ represents
the threshold separating long- and short-lived remnants.  The analysis presented in this work shows the potential of constraining it phenomenologically.

\item[(d)] {\em Short-lived Neutron Star} (\blue{SLNS}): If $M_{\rm ls}
\le M_{\rm tot} \le M_{\rm thr}$, where $M_{\rm thr}$ is the threshold
mass for prompt collapse to BH.
These are short-lived \blue{objects} that collapse on the GW timescale.

\item[(e)] {\em Prompt Collapse BH} (pcBH): If $M_{\rm tot}\ge
M_{\rm thr}$. The threshold mass is computed under the assumption that
the threshold mass is directly proportional to the TOV mass, 
\begin{equation}
M_{\rm
  th=}\,k_{\rm th}\, M_{\rm TOV},
  \label{eq:Mth}
\end{equation}  
  where the constant $k_{\rm th}$
  has
been evaluated using the fit provided in \cite{Kashyap2022}
for a variety of NS EoS.
\end{itemize}

We further account for a
dependence of $M_{\rm th}$ on the mass ratio $q$ using the
fitting formula provided by \citet{Perego2022} for the function
$f(q)= M_{\rm th}(q) /M_{\rm th}(q=1)$,
\begin{equation}
f(q) = \alpha(q) q + \beta(q) =
    \begin{cases}
    \alpha_l q + \beta_l & {\rm if}~q < \tilde{q} \, , \\
    \alpha_h q + \beta_h & {\rm if}~q \geq \tilde{q}    \, .
    \end{cases}
    \label{eq: double linear fit}
\end{equation}
where the parameter $\tilde{q}$,
approximately in the range $0.7-0.75$, separates two different regimes as evident from the compilation of a large number (250) of numerical BNS simulations. Following \cite{Perego2022}, we adopt the optimal value of $\tilde{q}=0.725$, and determine the parameters $\beta_l$ and $\beta_h$ in the equation above by imposing the condition $f(q=1)=1$, as well as the continuity of $f(q)$ at $q=\tilde{q}$. 

\blue{While using Eq.~\ref{eq:Mth} and \ref{eq: double linear fit} as part of our fiducial model, in the appendix we will quantitatively explore the effect of variations from this adopted fit. }

\subsection{Disk Mass Remnant}

We compute the disk mass according to the fitting formulae presented by \citet{Pang2024}, and based on  
\cite{Dietrich2020}.
These were derived by compiling the results from 73 general relativistic simulations of BNS mergers
by different groups (\cite{Hotokezaka2011,Dietrich2017,Radice2018,Kiuchi2019}), and using the same functional form derived by Coughlin {\em et al.} 2018 \cite{Coughlin2018} but with a correction to account for the dependence of the disk mass on the mass ratio $q$, 
\begin{eqnarray}
\log_{10}[M_{\rm disk}(M_\odot)] =  && \nonumber \\
{\rm max}\left\{-3, a\left(1 + b\tanh 
\left[\frac{c-(M_1+M_2)/M_{\rm thr}}{d}\right]\right) \right\}\,,
\label{eq:disk}
\end{eqnarray}
where the  parameters $a$ and $b$  are given by 
\begin{equation}
a=a_0+\delta a \cdot \Delta,\,\,\,\,\, b=b_0+\delta b \cdot \Delta\,,
\end{equation}
where
\begin{equation}
\Delta = \frac{1}{2}\tanh[\beta(q-q_{\rm trans})]\,.
\end{equation}
The various  best-fit parameters have the following numerical values:
$a_0=-1.581$, $\delta a=-2.439$, $b_0=-0.538$, $\delta_b = -0.406$, $c=0.953$,
$d=0.0417$,
$\beta=3.910$, $q_{\rm trans}=0.900$. 

\blue{This disk fit is a  component of our fiducial model. However, in the appendix we will explore how varying it in different ways impacts our results.
 }

\section{\label{sec:results} connecting BNS merger outcomes to NS nuclear properties}

\subsection{The post-merger \blue{object} population}

To connect the various elements laid out in the previous section, we perform Monte Carlo simulations to determine the \blue{frequency of the various outcomes and the corresponding disk mass distributions}  from BNS mergers. 

Given the initial NS mass distribution, we randomly draw the binary NS masses from it. Within our simple, \red{random pairing} assumption model, which assumes that the masses of the two NSs in the binary are uncorrelated with each other, both the primary mass $M_1$ and the secondary mass $M_2$ are drawn from the same distribution. 
Given two random drawings, the larger of the two masses is assigned to be the primary and the other is the secondary.

For any \blue{binary of total mass}
$M_{\rm
  tot}$, 
  we evaluate which of the five conditions listed in Sec.~IIB is satisfied, and assign the \blue{outcome}
  to the corresponding group. We perform this computation for a wide range of 52 EoSs taken from Table 2 of Ref.~\cite{Perego2022}, spanning the current range of nuclear physics uncertainties, for which we use $M_{\rm TOV}$ as a proxy to classification. 

\begin{figure}
    \centering
    \includegraphics[width=1.\linewidth]{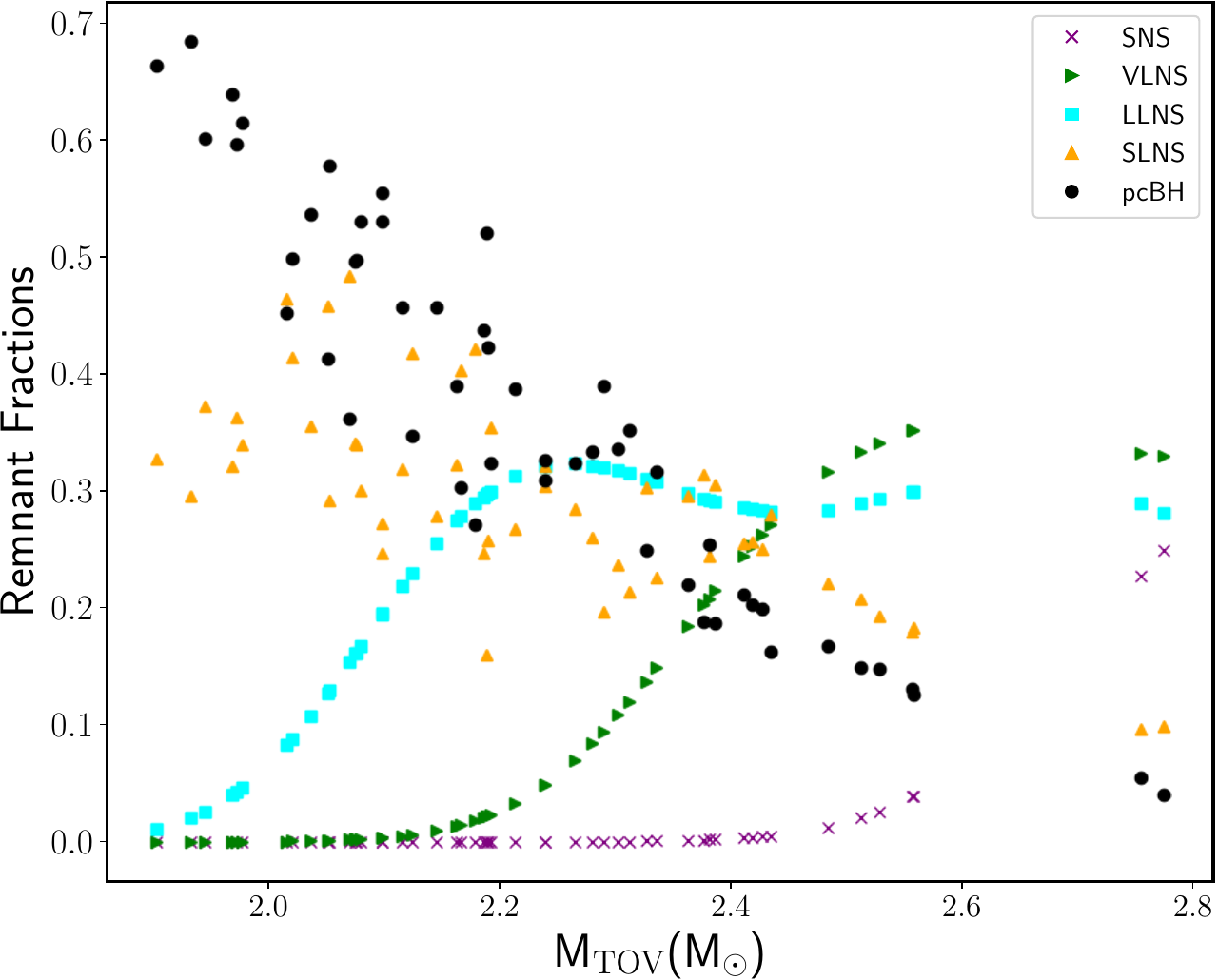}\\
    \caption{ Remnant fractions as a function of the total
      mass of the binary merger remnant,  where both
      $M_1$ and $M_2$ are independently drawn from the bimodal NS mass
      distribution. Here we adopted an intermediate, representative value $M_{\rm ls}=1.3 M_{\rm TOV}$. A larger (smaller) value would increase (decrease) the relative abundance of  \blue{LLNSs} with respect to \blue{SLNSs}.}
\label{fig:fractions}
\end{figure}

\begin{figure*}
    \centering
    \includegraphics[width=0.47\linewidth]{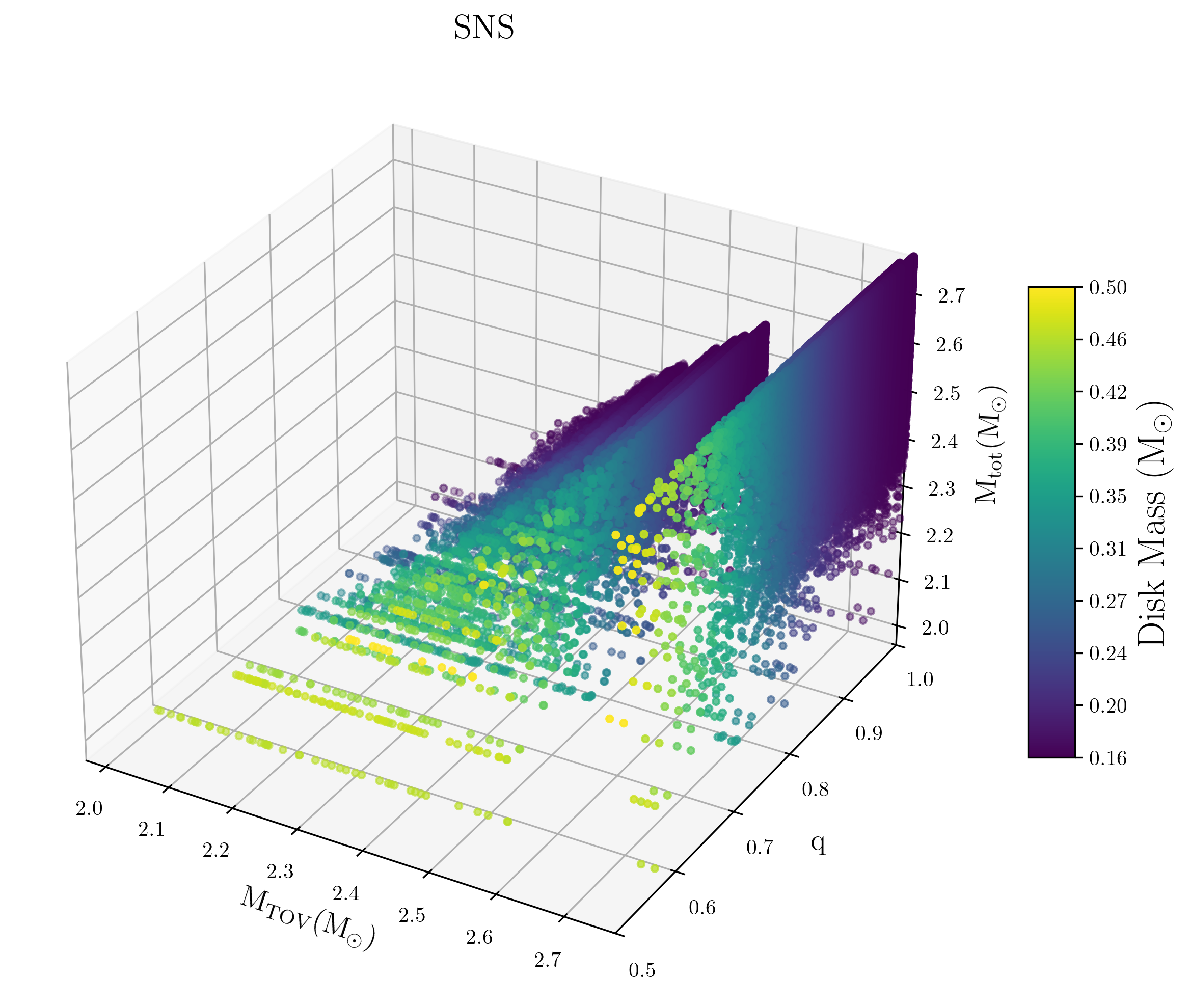}
    \includegraphics[width=0.47\linewidth]{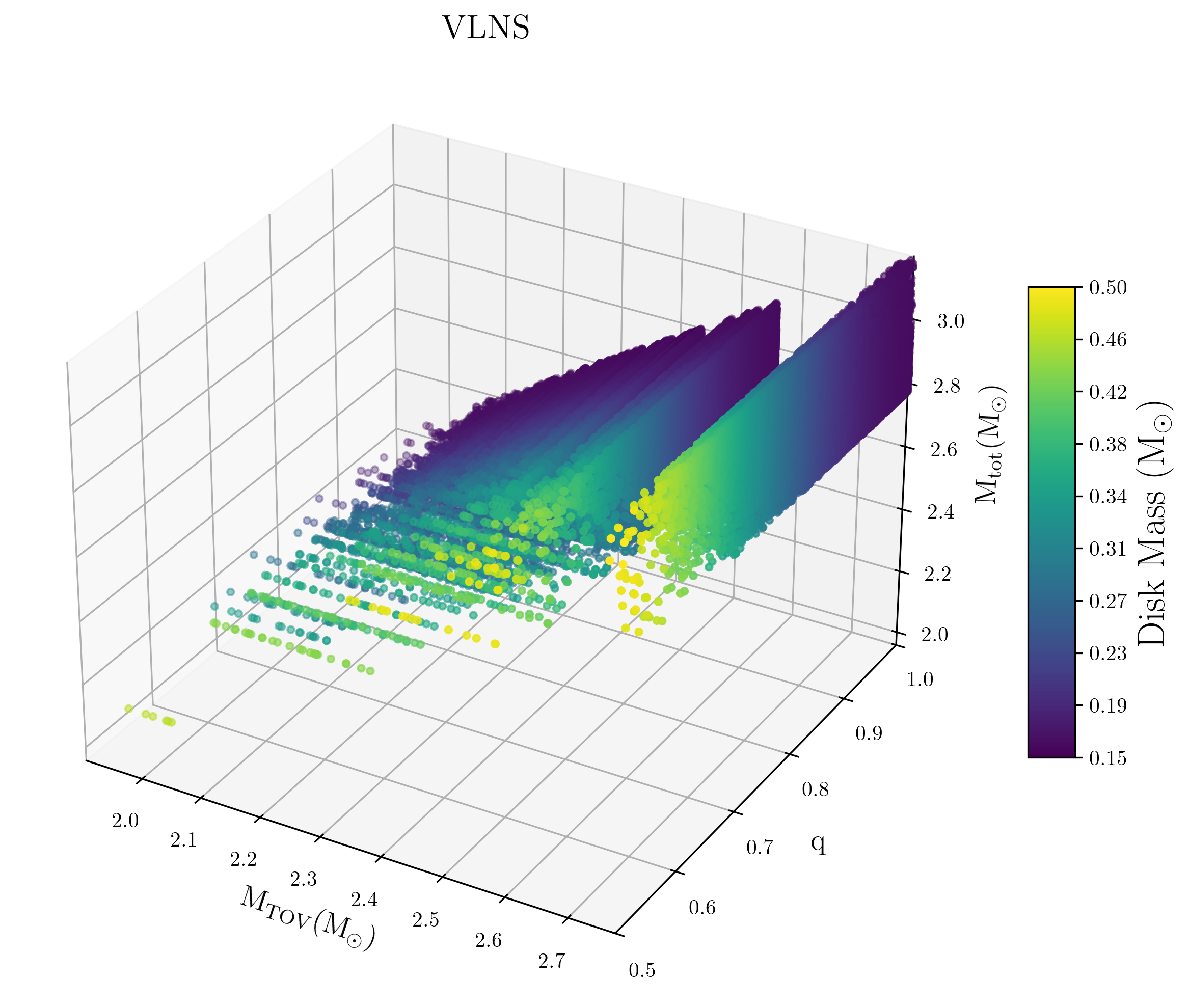}\\
    \includegraphics[width=0.47\linewidth]{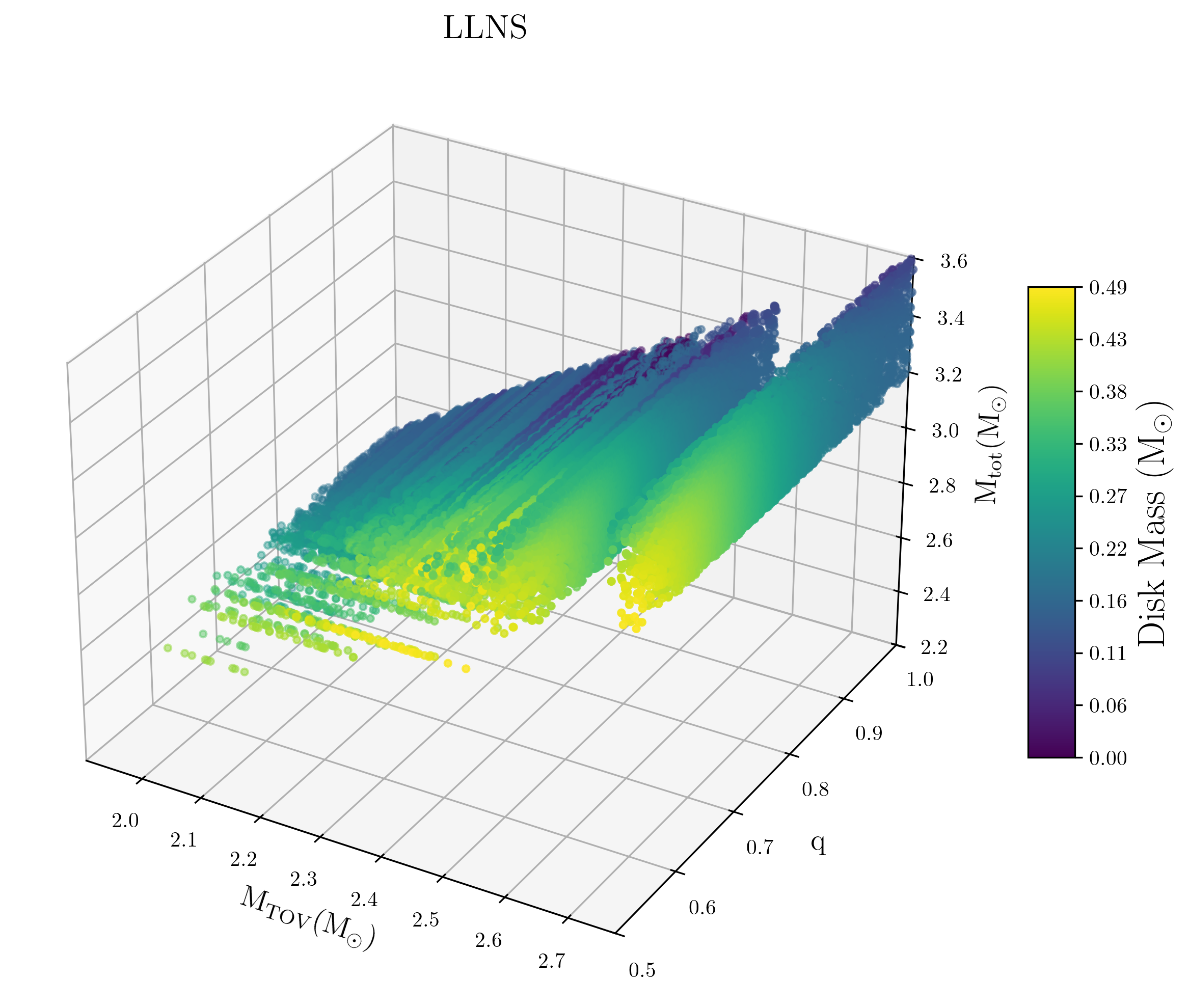}
    \includegraphics[width=0.47\linewidth]{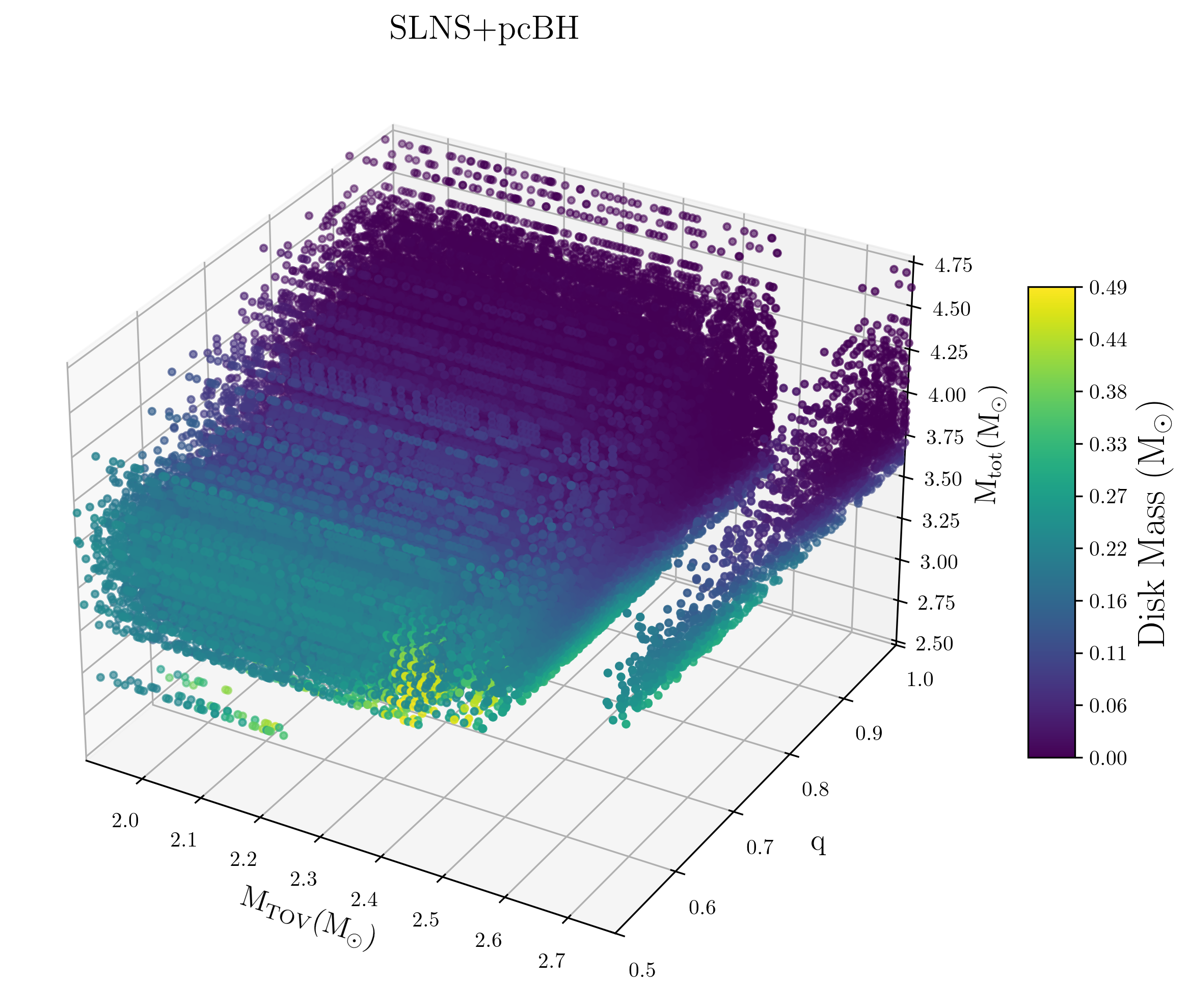}    
    \caption{Post-BNS merger disk mass as a function of the NS EoS (parametrized via $M_{\rm TOV}$), the mass ratio $q$ of the merging NSs, \blue{and the total merger mass $M_{\rm tot}$.} Here we used a representative value $M_{\rm ls}=1.3 M_\odot$ for the \blue{LLNS} to \blue{SLNS} transition. A larger value would shift some remnants from the \blue{SLNS}+pcBH sample to that of \blue{LLNS} in the bottom plots, and vice versa for a smaller value.}
\label{fig:disk}
\end{figure*}

Fig.~\ref{fig:fractions} shows the contribution of each  group, as a function of $M_{\rm TOV}$.  The overall trends are as expected: the larger $M_{\rm TOV}$, the higher the relative fraction of stable NSs, and the lower the fraction of prompt-collapse BHs. 
While the fractions of SNSs, \blue{VLNSs}, and \blue{LLNSs} follow a relatively smooth trend (other than for statistical fluctuations), the fractions of \blue{SLNSs} and pcBHs display relatively large fluctuations deviating from the overall trend. This stems from the fact that, while we have ordered the EoS by their $M_{\rm TOV}$, the corresponding values of pcBH do not strictly follow the same trends, i.e. some EoS may have a larger $M_{\rm TOV}$ but a smaller pcBH, hence resulting in non-strictly monotonic trends for the corresponding \blue{SLNS} and pcBH remnant fractions when plotted versus 
$M_{\rm TOV}$.

For each BNS merger in our MC simulations, we estimate the disk mass using the fit in Eq.~\ref{eq:disk}. We separate the mass distributions for each \blue{outcome} but combine the contributions of the \blue{SLNSs} and the pcBHs since they yield similar outcomes in our analysis. The disk masses are displayed as a function of 
$M_{\rm TOV}$, $q$ \blue {and $M_{\rm tot}$} in Fig.~\ref{fig:disk}. The dependence of the disk mass on the mass ratio is evident for all the \blue{cases}, but especially pronounced for the \blue{SLNS}+pcBH population, where the disk mass can vary from several tens of solar masses for high unequal mass ratios to negligible disk mass in the limit of equal mass ratios. Stable NSs have a generally large disk mass, from about  $\sim 0.5 M_\odot$ for a small mass ratio down to $\sim 0.15-0.2 M_\odot$ (with the precise value depending on the EoS) for equal mass ratios. The disk mass for the \blue{VLNS} population bears similarities to the stable NS 
while the \blue{LLNS} case is more similar to the \blue{SLNS}+pcBH population,
albeit with a gentler dependence on the mass ratio.
\blue{The scatter plots of Fig.~\ref{fig:disk} also illustrate the range of $M_{\rm tot}$, and the relative frequency, contributing to each type of merger outcome.}

\subsection{Long and Short GRBs from BNS mergers}

We connect the properties of the \blue{BNS merger outcomes}
 with the findings of \citet{Gottlieb2023} and the rate information to date of the lbGRB/sbGRB fraction. 
From the compilation of \citet{Fong2022}, this fraction in the $z<0.3$ sample can be estimated at about 40-50\%, using
GRB~060614 \cite{Gehrels2006}, GRB~211211A \cite{Troja2022} and GRB230307A \cite{Yang2024} to represent the  lbGRB share.
We note that this is a rough estimate and possibly subject to change with more future data.
A more comprehensive
analysis extending the sample to $z\lesssim 0.5$ suggests that the long-to-short ratio might be even as high as $\sim 1$ (A. Levan, private communication). 
However, despite current uncertainties in the data, our goal here is to show the constraining power for the nuclear properties of NSs of connecting theoretical ideas for the compact binary GRB phenomenology with the observed population. While in this work we derive constraints based on the current data, the analysis can be refined with a larger wealth of observations in the future.

\citet{Gottlieb2023,Gottlieb2024} developed a theoretical framework linking binary merger populations with observed GRB populations, attributing variations in GRB durations to differences in the merger remnant and the mass of the post-merger accretion disk around BHs. They demonstrated that the duration of a BH-powered GRB jet is governed not by the accretion timescale but by the time required for the disk to reach a magnetically-arrested (MAD) state \citep{Tchekhovskoy2011}. Due to the high compactness of the post-merger disk, magnetic flux is advected onto the BH in a short time, setting an approximately constant jet power. As the disk transitions to the MAD state, magnetic flux reconnects, and the jet power then follows the mass accretion rate as $ P \sim \dot{M} \sim t^{-2} $, powering the GRB extended emission.

\citet{Gottlieb2023} found that for massive post-merger accretion disks with $ M_d \gtrsim 0.1\,M_\odot $, the jet duration must be $ t \gtrsim 10\,{\rm s} $ to align with observed GRB luminosities; shorter durations would lead to a jet more powerful than what observations suggest. This implies that massive BH disks likely power lbGRBs. Based on several indications, including kilonova modeling, the magnetic field amplification and the bimodal distribution, \citet{Gottlieb2024} concluded that sbGRBs are powered by long-lived remnant NSs. Building on the unified framework that connects binary populations to GRBs and kilonovae, we consider the following three central engines:

\begin{enumerate}

    \item Millisecond  SNSs and \blue{VLNSs} are disfavoured as GRB progenitors since their immense rotational energy is notably absent in GRB radio afterglows 
\cite{Metzger2014,Horesh2016,Beniamini2021,Wang2024}
    and optical kilonova signals. Numerical simulations of BNS mergers also generally suggest that a hypermassive NS phase is likely necessary to launch a successful GRB jet (see e.g. \cite{Margalit2015,Ciolfi2018, Murguia2017}, but also note \cite{Kiuchi2024} for a successful outcome). While we adopt this as our fiducial model, however we also explore how our results vary if this assumption is released and magnetars associated with \blue{VLNSs} make up a sizeable fraction of cbGRBs.

    \item The outcome of mergers resulting in \blue{SLNSs} or prompt BH formation depends on the disk mass. When the disk mass is high, $ M_d \gtrsim 0.1\,M_\odot $, the BH will inevitably produce a long-lived jet, leading to a detectable lbGRB. However, if the disk mass is $ M_d \lesssim 10^{-2}\,M_\odot $, then given the magnetic flux on the BH, proportional to $ B^2 \sim M_d $, is reduced, the jet would be too faint to detect.

    \item \blue{LLNSs} power the standard sbGRBs, where the NS remnant lifetime roughly determines the GRB duration. We note that \blue{LLNSs} are predicted to form massive accretion disks \citep[e.g.,][]{Radice2018}. This implies that in some cases, the collapse of the \blue{LLNSs} into a BH could extend the emission, potentially producing an lbGRB. However, the classification of the GRB would depend on the relative power of the NS-driven and BH-driven jets \citep[see discussion in][]{Gottlieb2024}. While some \blue{LLNSs} may contribute to the lbGRB population, for the following analysis we assume that all \blue{LLNSs} contribute exclusively to the sbGRB population.
    
\end{enumerate}

To match the observed GRB populations, the rate of \blue{LLNSs} must be comparable to that of \blue{SLNSs} and pcBHs with massive disks, which produce detectable lbGRBs.

\begin{figure}
    \centering    \includegraphics[width=1.0\linewidth]{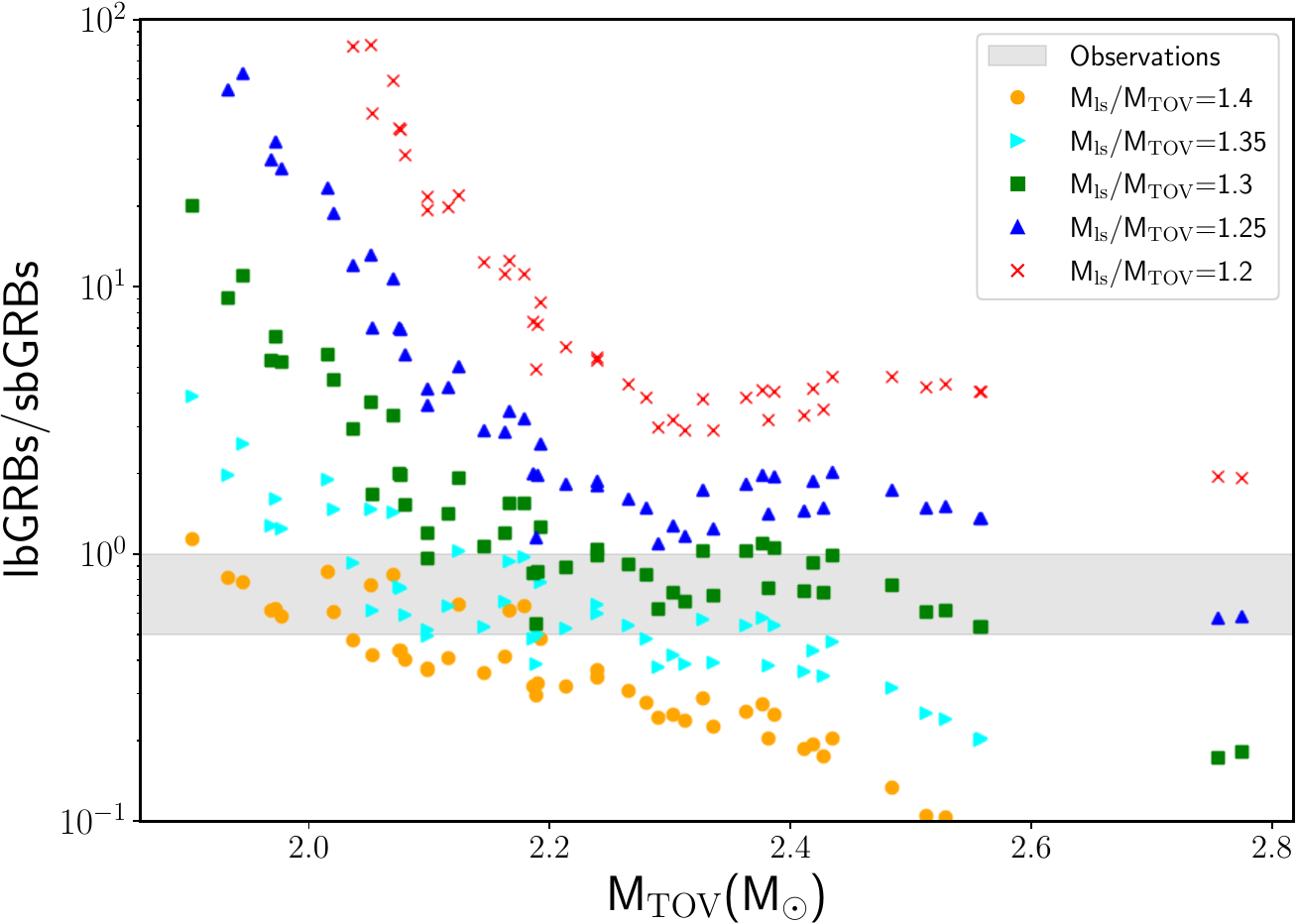}
    \caption{The ratio between sbGRBs and lbGRBs, where the former are identified with the \blue{LLNSs} and the latter with the combined populations of \blue{SLNSs} and pcBH with a disk mass larger than
    a minimum value. Five different limits are considered for the transition mass $M_{\rm ls}$ between a long-lived and a short-lived NS remnant.
    The grey band denotes the current observational constraints.}
\label{fig:GRBs}
\end{figure}

 Fig.~\ref{fig:GRBs} depicts the ratio of \blue{LLNSs} to (\blue{SLNSs} + pcBHs) with disk mass $\geq 0.1 M_\odot$ as a function of $M_{\rm TOV}$, and for five different choices of the transition mass $M_{\rm ls}$ in a broad range of values between
 1.2-1.4~$M_{\rm TOV}$. \blue{We should however note that for 8/52 of the EoS explored here\footnote{\blue{These are: APR ($M_{\rm TOV=}2.188M_\odot$), GMSR(H1) ($M_{\rm TOV=}2.289M_\odot$), GMSR(H3) ($M_{\rm TOV=}2.302M_\odot$), 
 GMSR(H4) ($M_{\rm TOV=}2.336M_\odot$), GMSR(H5) ($M_{\rm TOV=}2.381M_\odot$),
 GMSR(DHSL69) ($M_{\rm TOV=}2.411M_\odot$), 
 GMSR(DHSL59) ($M_{\rm TOV=}2.427M_\odot$),
 GMSR(H7) ($M_{\rm TOV=}2.512M_\odot$).
 }}, the condition $1.4M_{\rm TOV}<M_{\rm th}$ is not satisfied. While this does not affect the computed fractions (in this case the SLNS+pcBH population simply reduces to the pure pcBH in the Monte Carlo), however we should note that for those EoS the LLNS-SLNS transition must occur for
 $a_{\rm crit}< 1.4$ and therefore the corresponding points for $a_{\rm crit}= 1.4$ should not be interpreted as a constraint on this parameter. In those specific cases, the lbGRBs/sbGRBs ratio sensitivity is rather shifted to $M_{\rm th}$.}
 The grey band in the figure guides the eye to the current estimated rates, for which
we have adopted a broad conservative range of $50\% -100\%$.
Taking these rates at face value, $M_{\rm ls}=1.2 M_{\rm TOV}$ is largely inconsistent with the data, for any choice of the NS EoS. 
A transition at $M_{\rm ls}=1.25 M_{\rm TOV}$ would be compatible with the data but only for extremely stiff EoS, with 
$M_{\rm TOV}\gtrsim 2.7 M_{\odot}$, which are  disfavoured by GW170817  observations \cite{Margalit2017,Rezzolla2018,Shibata2019}. On the other hand, a transition around the higher value $M_{\rm ls}=1.3 M_{\rm TOV}$ is consistent with the data for a range of acceptable EoS
with $M_{\rm TOV}\lesssim 2.6 M_\odot$. Similarly, a transition
at $M_{\rm ls}=1.35 M_{\rm TOV}$
is acceptable within a broad range of viable EoSs with $2.1 M_\odot\lesssim M_{\rm TOV}\lesssim 2.4 M_\odot$,
while the very high value 
$M_{\rm ls}=1.4 M_{\rm TOV}$ can be accommodated, but only within the softer EoS range
with $M_{\rm TOV}\lesssim 2.2 M_\odot$.
Therefore, we conclude that the transition between the long- and short-lived NS remnants likely occurs in the range $\sim [1.3-1.4]M_{\rm TOV}$.

 We note that a premise of our analysis is that
 the contribution to lbGRBs from NS--BH mergers is negligible compared to NS--NS mergers. 
 This is motivated by the fact that numerical relativity simulations of NS--BH mergers \cite{Foucart2012} show, for a variety of mass ratios and BH spin, that the disk mass is generally small. Massive disks can only be obtained for less massive BHs or very rapidly spinning BHs\footnote{Both NS--BH events detected to date are in the regime of high-mass ratio and low spin \citep{Abbott2021NSBH}. Thus, if representative of the population, detectable EM counterparts are likely to be uncommon.}. Hence, even if NS--BH mergers were to contribute to lbGRBs, they would likely be too dim to detect. \blue{However, more generally, note that any contamination to the observed sample of long cbGRBs by NS-BH mergers would mean that the fraction of long cbGRBs from NS-NS is smaller, hence strengthening our results.}

 Our analysis assumes that the \blue{VLNS} population does not contribute to a sizeable fraction of cbGRBs for the reasons discussed earlier. However, as noted by \citet{Margalit2022}, some initially long-lived magnetars may evolve similarly to \blue{LLNSs}, potentially powering an sbGRB.
 Therefore, we assess the implications of this scenario on our results by performing two additional simulations that consider the two extreme cases of the \blue{VLNS} population contributing fully to either the short or the long cbGRB components, depending on whether the \blue{VLNSs} collapse on a timescale of $ \lesssim 1\,{\rm s} $ or $ \sim 10\,{\rm s} $, respectively.
We find that if \blue{VLNSs} contribute exclusively to sbGRBs, the transition with
$a=1.2$ is still inconsistent with the data for $M_{\rm TOV} <  2.5 M_\odot$,  while the $a=1.25$ threshold is only allowed for $M_{\rm TOV} \gtrsim  2.4 M_\odot$. 
The higher value $a=1.3$ has the best consistency in the 2.2-2.4~$M_\odot$ TOV range of masses, while higher values of $a$ would require softer EoSs closer to 2.2~$M_\odot$. Conversely, if \blue{VLNS} contribute to the long cbGRB population only, we find that,
within the 2.2-2.4~$M_\odot$ range of TOV masses,
$a\lesssim 1.3$ is borderline consistent with the data, while both $a=1.35$ and $a=1.4$ 
have a broad range of consistency. These cases discussed above would constitute the most extreme departures from our fiducial model since either a smaller contribution from the \blue{VLNS} population or a mixed contribution to both the long and short cbGRBs would make the final estimates closer to those shown in Fig.~\ref{fig:GRBs}. 
Therefore, our results of a relatively high transition mass $M_{\rm ls}\gtrsim 1.3 M_{\rm TOV}$
are quite robust within the limits and assumptions of the unified GRB model and the available observational data to date. \blue{A broad quantitative assessment of the robustness of our results with respect to the various model inputs is provided in the Appendix.}

\subsection{Comparison with numerical relativity}
An important result of our analysis is the demarcation point separating short- and long-lived remnants from BNS mergers at $\gtrsim 1.3 M_{\rm TOV}$. This is somewhat higher than the typically quoted ``supramassive'' threshold $\simeq 1.2 M_{\rm TOV}$ \cite{Baumgarte2000, Lasky2014, Breu2016, Piro2017}, where supramassive remnants are likely very long-lived according to the classification used here. As we show here, this result is consistent with current numerical-relativity simulations. In particular, we use data from 273 NS merger simulations from the \texttt{CoRe} catalog \cite{Dietrich:2018phi, Gonzalez:2022mgo}. This sample is essentially the entire catalog, with the exclusion of 24 binaries whose waveforms show unphysical artifacts. The data spans 12 EOSs, total mass between $2.4\, M_\odot$ and $3.4379\, M_\odot$, mass ratio $0.4856 \leq q \leq 1$. $M_{\rm TOV}$ varies between $1.7833 M_\odot$ and $2.8349 M_\odot$ among the EOSs we consider. Some of the simulations also consider spinning and/or eccentric binaries \cite{Gonzalez:2022mgo}. The precise time to collapse for BNS remnants is notoriously sensitive to numerical details and cannot be extracted reliably from simulations \cite{Zappa:2022rpd}. However, well-resolved simulations can robustly distinguish between short- and long-lived remnants. Moreover, our analysis considers simulations performed with two independent codes, \texttt{BAM} \cite{Bruegmann:2006ulg, Thierfelder:2011yi} and \texttt{THC} \cite{Radice:2012cu, Radice:2013xpa, Radice:2015nva}. We only discuss results that are robust and consistent between the two codes and across the \texttt{CoRe} database.

\begin{figure}
    \centering
    \includegraphics[width=1.0\linewidth]{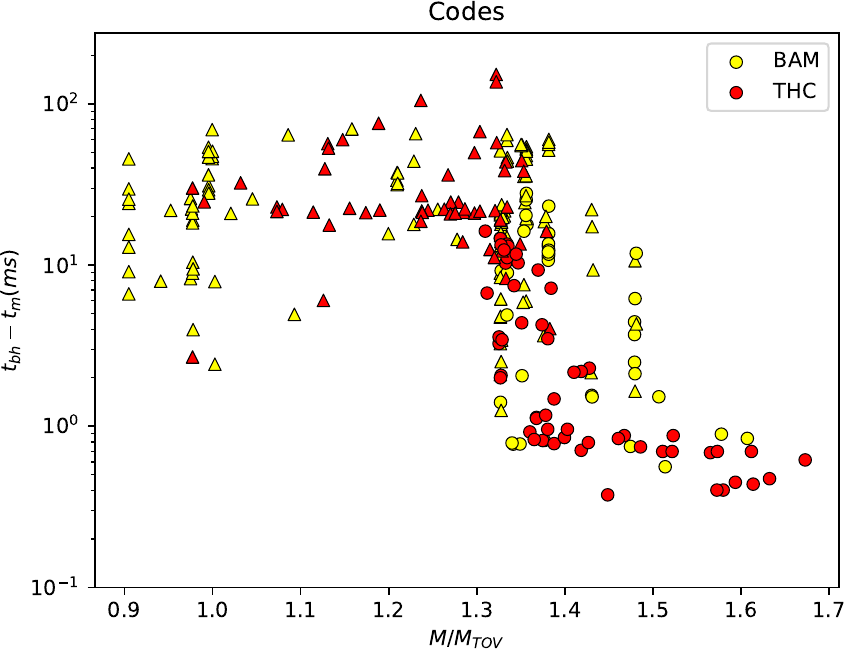}
    \caption{Time delay between merger and BH formation for 273 numerical relativity simulations. Lower limits are reported as triangles, while the simulation results are indicated with circles.}
    \label{fig:nrdata}
\end{figure}

As customary, we define the time of merger $t_{\rm merg}$ as the time at which the amplitude of the $\ell=2,m=2$ multipolar component of the gravitational-wave strain peaks. Subsequently, we find candidates for the time of BH formation using the \texttt{find\_peaks} function of \texttt{scipy} on the $\ell=2,m=2$ strain amplitude. We validate and, where necessary, manually adjust $t_{\rm BH}$ by visually inspecting all of the waveforms. If BH formation does not happen during the simulation time, we report a lower limit given by the time up to which each simulation was performed.

Our results are summarized in Fig.~\ref{fig:nrdata}. The data shows clear evidence for a transition around ${\sim}1.3 \blue{M_{\rm TOV}}$, consistent with the population synthesis results. The transition region is broad, spanning $1.3 \lesssim \blue{M_{\rm ls}}/M_{\rm TOV} \lesssim 1.4$, suggesting that there is additional physics determining the outcome of the merger, in addition to $M_{\rm TOV}$. For example, the mass ratio $q$ is known to play an important role in setting the angular momentum of the remnant \cite{Bernuzzi:2014kca}. Moreover, there might be peculiar features in the EOSs that influence the outcome of mergers, beyond $M_{\rm TOV}$. A more accurate determination of the separation between long- and short-lived remnants is beyond the scope of this work. For example, it would be interesting to reformulate the criterion for the formation of short- vs long-lived remnants probabilistically and to explore its crucial physical dependencies. In combination with estimates for $M_{\rm TOV}$, this would have the potential to provide novel constraints on the EOS of dense matter in the most extreme conditions reached in mergers. Unfortunately, the binaries included in the \texttt{CoRe} database were not selected randomly, and, in particular, some binaries are overrepresented (e.g., GW170817-like configurations), so our current numerical relativity data is not yet sufficient to realize this goal.

\section{\label{sec:discussion}summary and discussion}

We proposed a novel connection between the phenomenology of GRBs from binary mergers and the nuclear properties of NSs. In particular, within the theoretical framework of \citet{Gottlieb2023,Gottlieb2024}, we have shown that the relative fraction of long and short cbGRBs has a very sensitive dependence on the transition mass $M_{\rm ls}=a M_{\rm TOV}$ between long-lived and short-lived NS remnants. 

We find that a relatively large value for the transition between long and short-lived NS remnants,  $M_{\rm ls}\gtrsim 1.3 M_{\rm TOV}$, is strongly favoured by the current data for a range of viable EoSs.
This implies that the fraction of mergers resulting in the formation of long-lived remnants might extend over a broader mass range than traditionally expected. 
\red{We remind the reader that our modeling relies on various input parameters, some of which are better constrained than others. While we adopted a fiducial model based on the best fits to current observational and numerical data from the literature, we have explored in the appendix how our conclusions would change if these input parameters deviated from their assumed values. Of particular importance is the dependence on the NS mass function for both the primary and secondary, which we expect to be more robustly calibrated in the future as GW and radio data of binary NSs grow. }

Motivated by our phenomenological results, we then searched for an indication of the long-short-lived transition in a subset of numerical simulations robust enough to resolve such regime change. While the transition region is found to be broad in the numerical data, there is clear evidence for a transition at $\sim 1.3M_{\rm TOV}$, in agreement with our findings.

Our results disfavour scenarios, such as certain types of QCD phase transitions, pion or kaon condensation, or other exotic physics that would lead to catastrophic pressure loss at several times nuclear density and temperatures of tens of MeV leading to the rapid gravitational collapse of binaries with total mass ${\color{blue}M_{\rm tot}} \lesssim 1.3 M_{\rm TOV}$ \cite{Margalit2022}. That said, the formation of short-lived remnants with ${\color{blue}M_{\rm tot}} \gtrsim 1.3 M_{\rm TOV}$ can still be accommodated if $M_{\rm TOV} \gtrsim 2.6 M_\odot$, which however is in tension with constraints from GW170817 \cite{Margalit2017, Shibata2019}
\blue{and with multi-messenger NS data inference by \cite{Fan2024}.}
Our findings are consistent with the interpretation of the blue kilonova in GW170817 as a spiral-wave driven wind \cite{Nedora2019, Nedora2022, Bernuzzi2024}, or as the magnetized wind from a long-lived, but ultimately unstable magnetar \cite{Metzger2018, Combi2023}.

While the analysis presented here has been of statistical nature and has only relied on the relative fraction of long and short GRBs from BNS mergers to constrain $M_{\rm ls}$, future observations of GWs with GRBs have the power to further bound this threshold mass. In particular, if the total mass $M_1 + M_2$ is accurately measured from GW observations, and the disk mass is also observationally estimated from a kilonova detection, then whether the accompanying cbGRB is found to be short or long can allow to bound the value of  $M_{\rm ls}$ with respect to the remnant mass.

\begin{acknowledgements}
We thank Alejandra Gonzalez, Patrick Bush III, and Rossella Gamba for help with the \texttt{CoRe} database data;  Floor Broekgaarden, Will Farr, Michela Mapelli, Soumendra Roy and Lieke van Son for informative discussions on the neutron star mass distribution, and Brian Metzger for very useful comments on our manuscript. 
We are especially grateful
to Jillian Rastinejad and Andrew Levan for sharing their pre-publication data on the relative rates of long and short cbGRBs. \blue{Last we thank an anonymous referee for insightful and constructive comments which helped to strengthen our work.}
RP acknowledges support from the National Science Foundation under Grant AST-2006839.
OG is supported by the Flatiron Research Fellowship. The Flatiron Institute is supported by the Simons Foundation. 
ES was partially supported by NASA through Award Number 80NSSC21K1720.
DR acknowledges support from the Sloan Foundation, from the U.S.~Department of Energy, Office of Science, Division of Nuclear Physics under Award Number(s) DE-SC0021177 and DE-SC0024388, and from the National Science Foundation under Grants No.~PHY-2020275, AST-2108467, PHY-2116686, and PHY-2407681.

\end{acknowledgements}

\appendix*
\blue{
\section*{{Appendix: Dependence of results on the relevant input parameters} }
}

\blue{
Here, we quantitatively examine how our predicted lbGRBs/sbGRBs ratio depends on the main input parameters. We vary one parameter at a time, exploring deviations from the fiducial model in each case. Since an increase or decrease in the lbGRBs/sbGRBs ratio causes all the curves in Fig.~\ref{fig:GRBs} to shift accordingly, we focus our analysis on a single reference value of \( M_{\rm ls}/M_{\rm TOV} \), specifically \( 1.3 \). 
}
\bigskip

\begin{figure}
    \centering    \includegraphics[width=1.0\linewidth]{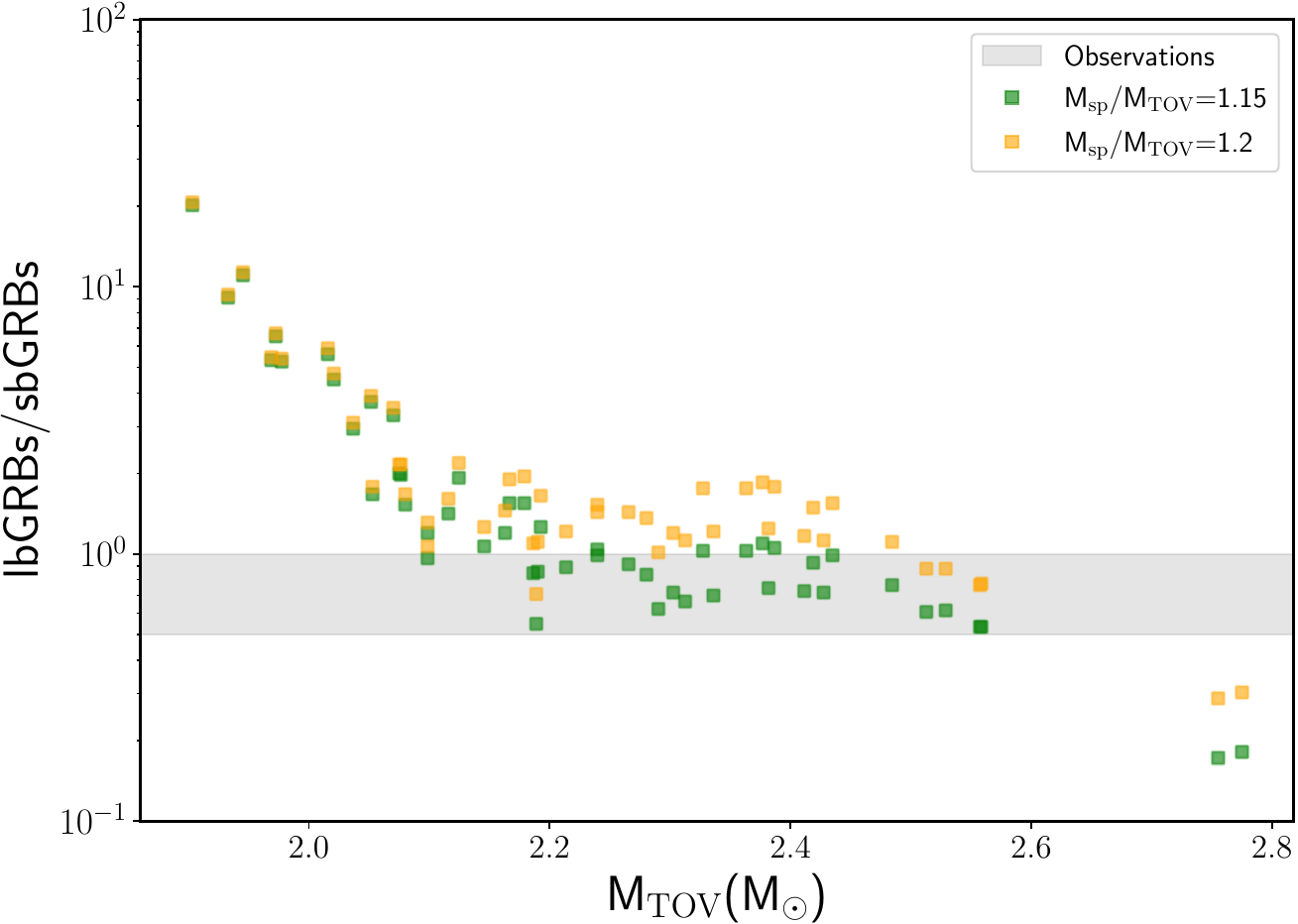}
    \caption{\blue{Dependence of the ratio between sbGRBs and lbGRBs on the transition mass from VLNSs to LLNSs. The choice of $M_{\rm sp}/M_{\rm TOV}=1.15$ adopted here makes our results more robust with respect to the 1.2 value commonly adopted in the literature.}}
\label{fig:test_sp}
\end{figure}

\noindent
\blue{\em (a) Dependence on $M_{\rm sp}$}\\
\blue{Fig.~\ref{fig:test_sp} compares the ratio 
lbGRBs/sbGRBs between our fiducial, low-limit case of $M_{\rm sp}/M_{\rm TOV}=1.15$, with the higher value 
$M_{\rm sp}/M_{\rm TOV}=1.2$ suggested in the literature \cite{Breu2016}.
The figure clearly shows that our fiducial choice of $M_{\rm sp}/M_{\rm TOV}=1.15$ is conservative with respect to our results: a higher value of the maximum mass of an uniformly rotating NS would strengthen the case of a high transition mass $M_{\rm ls}$. }

\bigskip

\begin{figure}
    \centering    \includegraphics[width=1.0\linewidth]{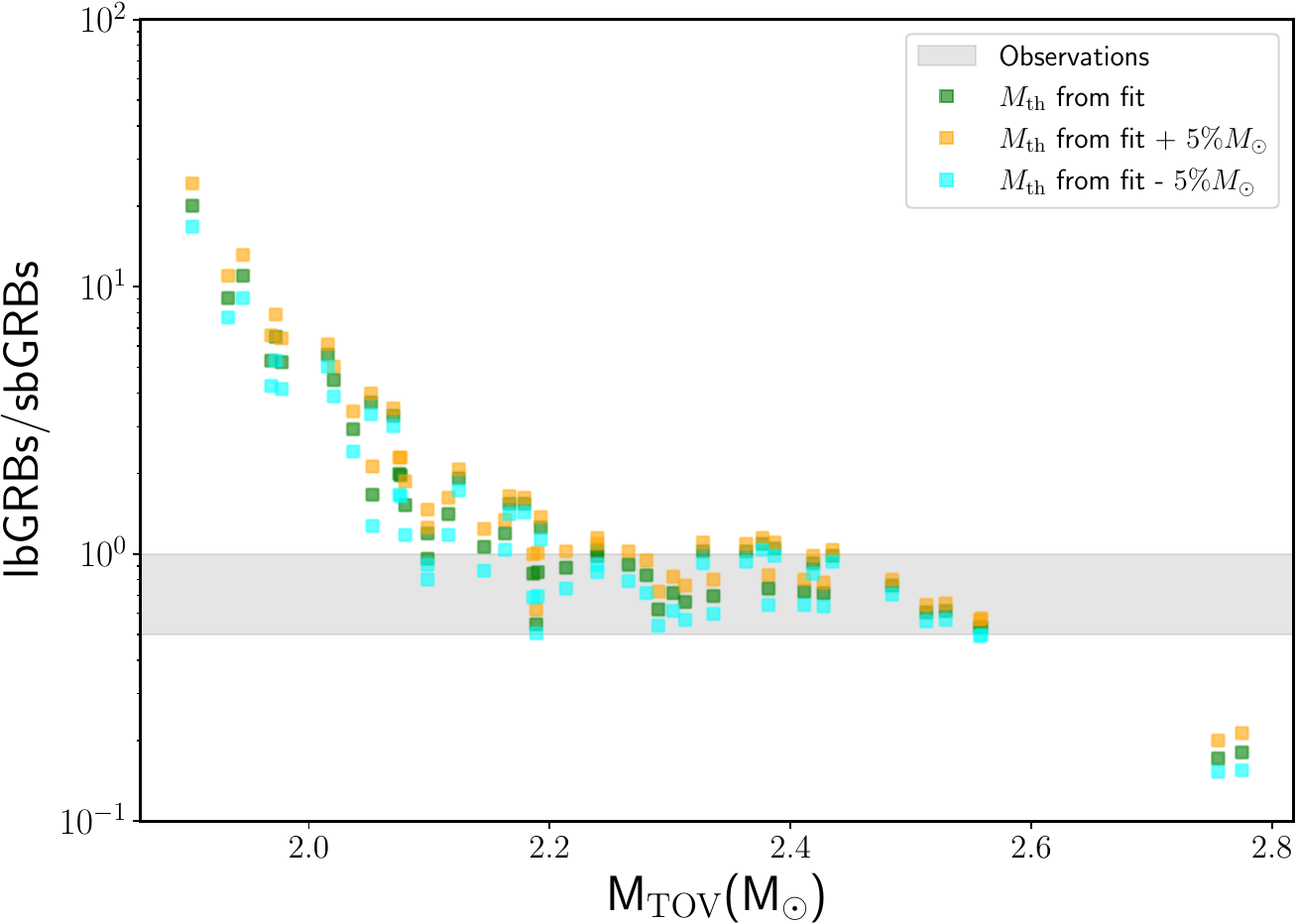}
    \caption{\blue{Dependence of the ratio between sbGRBs and lbGRBs on the threshold mass for prompt collapse to a BH. A 5\% $M_\odot$ variation around our adopted best fit to $M_{\rm thr}$ is considered, since various prescriptions from the literature generally agree within this range.}}
\label{fig:test_Mth}
\end{figure}

\noindent
\blue{\em (b) Dependence on $M_{\rm th}$}\\
\blue {While here we used the fit to $M_{\rm th}$ by \citet{Kashyap2022} and \citet{Perego2022}, however several other fits exist in the literature (e.g. \cite{Bauswein2021,Tootle2021,Kolsh2022}),
 generally agreeing within a few percent of a solar mass. 
 To evaluate the impact of variations in \( M_{\rm th} \) on our results, we performed two additional sets of simulations: one where \( M_{\rm th} \) was increased by a conservative \( 5\% \) of \( M_\odot \) and another where it was decreased by the same amount. 
  The small dependence on $M_{\rm th}$ derives from the presence of this variable in the expression for the disk mass fit. 
  Figure~\ref{fig:test_Mth} presents the results of this analysis, demonstrating that the variations induced by potential uncertainties in \( M_{\rm th} \) are minor compared to the trends observed with varying \( M_{\rm ls} \) in Figure~\ref{fig:GRBs}.
}

\bigskip

\noindent
\blue{\em (c) Dependence on $M_{\rm d}$}\\
\blue{As discussed in IIC, the disk mass adopted here is based on a fit produced from the combination of the results of a large number of numerical simulations. However, these are subject to various uncertainties, both of numerical nature as well as due to incomplete microphysics. Since these uncertainties are hard to exactly quantify, here we adopt an inverse approach: we study by what amount the fit to the disk mass needs to be incorrect for our results to weaken. }

\begin{figure}
    \centering    \includegraphics[width=1.0\linewidth]{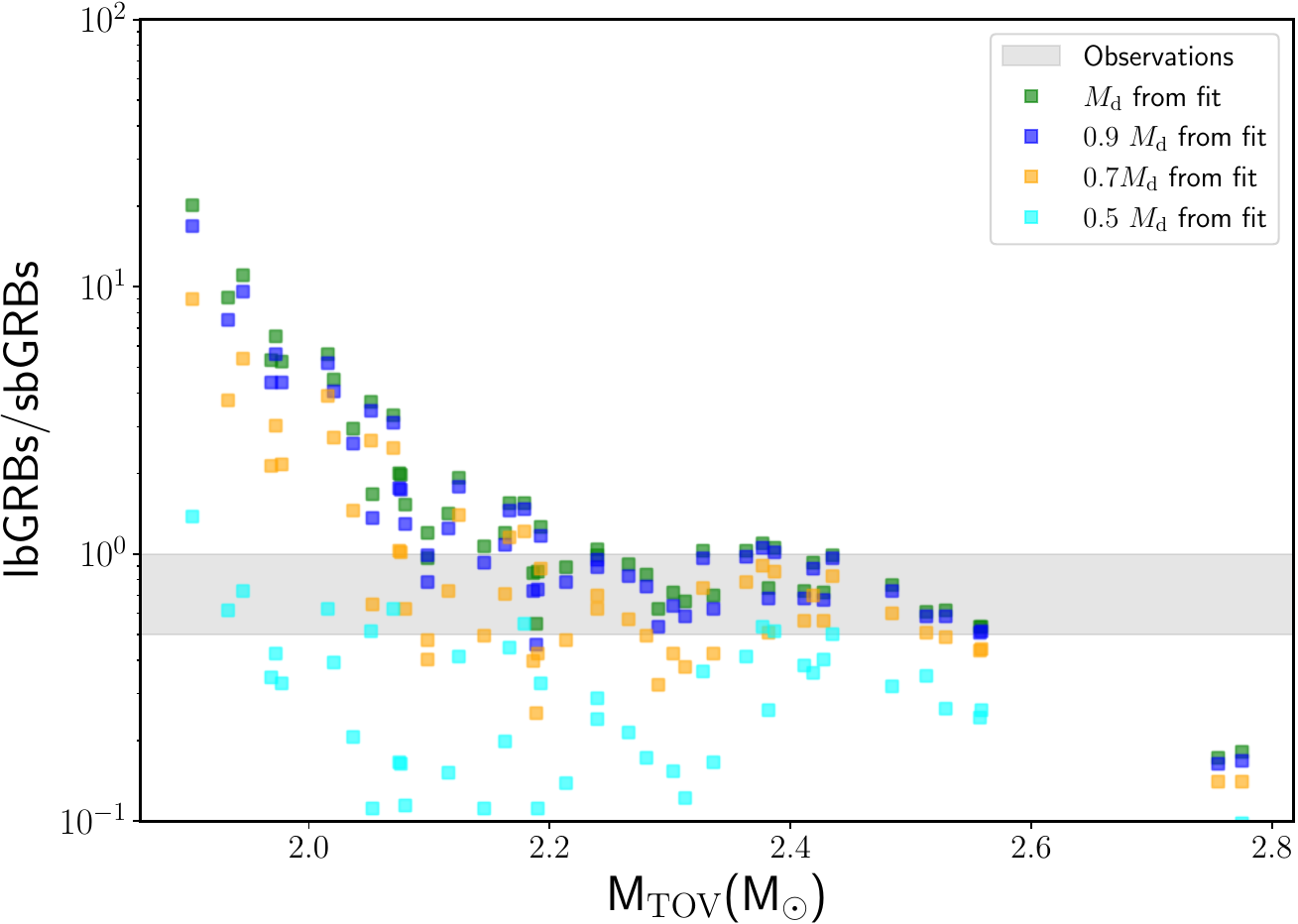}
    \caption{\blue{Dependence of the ratio between sbGRBs and lbGRBs on the mass of the disk around the post-merger object, making the assumption that the disk mass is systematically lower than what provided by the best fit, by various amounts. Higher values of the disk mass would increase the ratio and make our conclusions more robust.}}
\label{fig:test_Md}
\end{figure}

\blue {First we note that, if the disk mass was systematically underestimated, then our results would be even more robust towards high values of 
$M_{\rm ls}$, since the lbGRBs/sbGRBs ratio would shift upwards. 
Therefore, we present numerical results only for the scenario in which the disk mass is systematically overestimated by the fit.
Figure~\ref{fig:test_Md} illustrates how the ratio for the \( M_{\rm ls} = 1.3 \) case would change if the disk mass were \( 90\% \), \( 70\% \), or \( 50\% \) of the fitted value. As shown in the figure, for our conclusions to be significantly weakened, the disk mass would need to have been overestimated in the current fits by a factor of about 2.
 }

\bigskip

\noindent
\blue{\em (d) Dependence on the NS mass distribution}\\
\blue{Last we perform further simulations to explore possible deviations from the assumed NS mass function. This exploration is motivated by the fact that observational NS samples are heterogeneous, and could be biased by selection effects (see e.g. \cite{Chattopadhyay2020}). Ideally, the NS mass function (NSMF) as input for this analysis would be one calibrated on the subset of NS-NS systems which merge in the Hubble time (sample gathered from a combination from GW observations and radio surveys). However, until this subset is large enough, we can only work with the full sample, with the understanding that it may or may not be a faithful representation of the merging one. }

\begin{figure}[h]
    \centering    \includegraphics[width=1.0\linewidth]{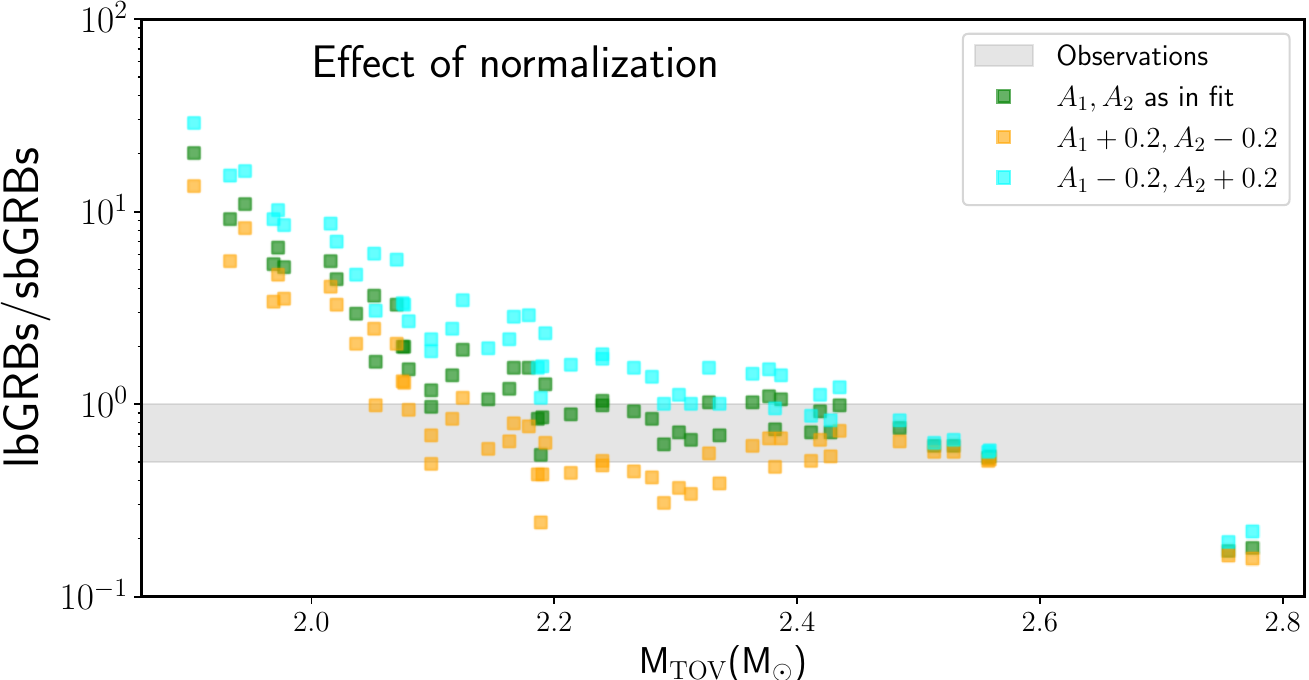}\\
    \includegraphics[width=1.0\linewidth]{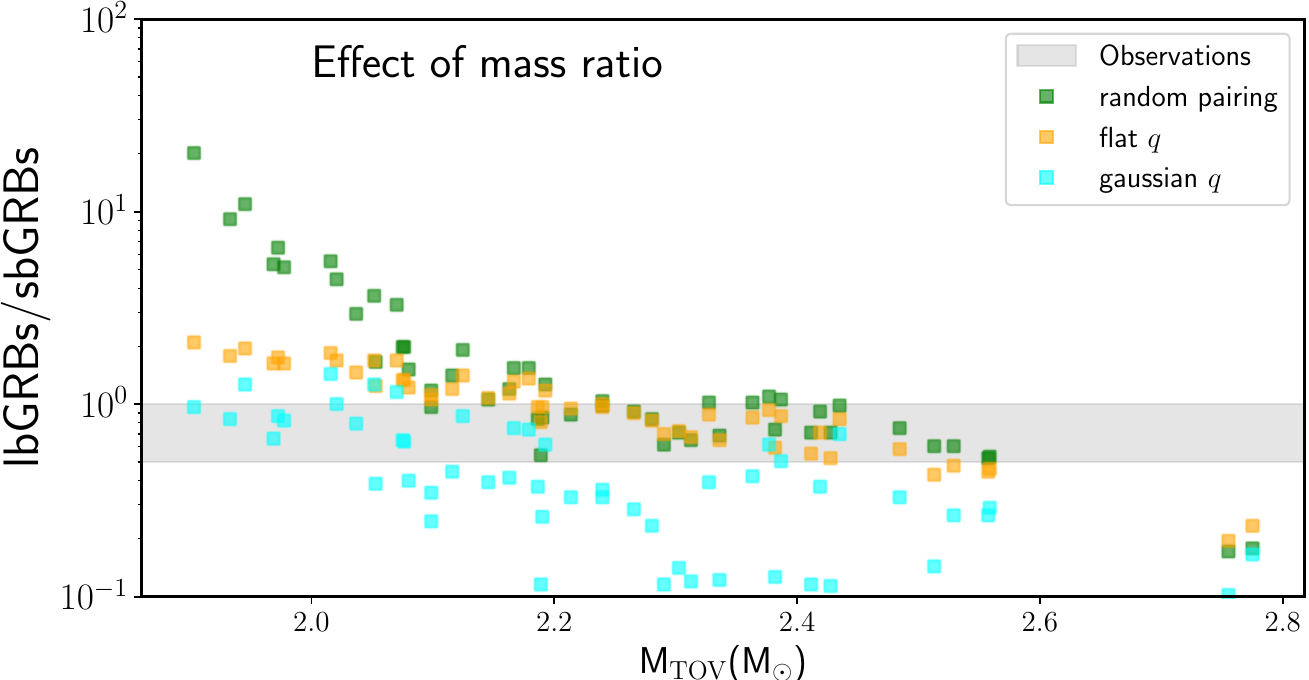}
    \caption{\blue{Dependence of the ratio between sbGRBs and lbGRBs on variations in the NS mass distribution wit respect to our adopted best fit. The top panel explores the dependence on the relative strengths $A_1$ and $A_2$ of the low- and high-mass peaks of the NS mass distribution, respectively. In the lower panel the NS mass distribution of the primary is taken from the best fit, but, for exploratory purposes, the distribution of the mass ratio $q$ is assumed either flat between 0.5 and 1, or a relatively narrow one-sided Gaussian with maximum at 1 and dispersion $0.15$. }}
\label{fig:test_NSM}
\end{figure}

\blue{
For this analysis, we continue to adopt a bimodal NSMF, since this has been a robust finding with the latest NS samples by several independent researchers (e.g., \cite{Schwab2010, Antoniadis2016, Alsing2018, Farr2020, Galaudage2021, Rocha2023}), and it has been suggested to hold independently also for the GW subsample \cite{Galaudage2021}.  
However, we explore the impact of varying the relative fraction of low-mass and high-mass NSs on our results. To this end, we conducted several simulations, gradually adjusting the normalizations $A_1$ and \(A_2\) of the two peaks—from their fiducial values of 0.54 and 0.46, respectively—until noticeable deviations from the fiducial model emerged.}  

\blue{
The results are presented in the top panel of Fig.~\ref{fig:test_NSM}, where the normalizations of the two NS populations are each increased or decreased by a factor of 0.2, corresponding to a variation of approximately \( \sim 40\% \) relative to the fiducial values. If the high-mass population were more dominant, our conclusions would be reinforced, as the fractions of SLNSs and pcBHs would increase. Conversely, a relatively larger fraction of lower-mass NSs would reduce the occurrence of these remnant types. However, as shown in the figure, a substantial deviation from the fiducial (best-fit) values is required before our conclusions begin to weaken.  
 }

\blue{
Lastly, we examine the impact of the mass ratio $q$. In our fiducial model, we adopted a \red{random pairing} assumption, generating both
the primary and secondary masses from the same initial mass function. This resulted in a distribution peaked around $ q = 1 $
but with a broad tail. To explore how variations in $ q $ affect the
results, we consider two extreme models: one in which $ q $ is drawn from a flat distribution, and another in which it follows a Gaussian
distribution centered at $ q = 1 $ with a standard deviation of $\sigma = 0.15 $, which significantly favors systems with comparable
masses relative to the fiducial model.}

\blue{
The effect of $ q $ in our simulations is twofold. For a given primary mass distribution, lower values of $ q $ correspond to lower
secondary masses, leading to a decrease in $ M_{\rm tot} $. This, in turn, reduces the fraction of SLNS and pcBH remnants, thereby lowering the lbGRB/sbGRB ratio. Conversely, lower values of $ q $ tend to produce higher disk masses, which counteracts the impact on the
lbGRB/sbGRB ratio. Both of these effects are highly sensitive to the EoS, leading to non-monotonic trends in their influence on the results.
For the flat $q$ distribution, these opposing effects roughly balance each other, except at low values of $ M_{\rm TOV} $, where
the system is more sensitive to reductions in $ M_{\rm tot} $. On the other hand, a $ q $ distribution that strongly favors nearly
equal-mass systems would be primarily influenced by the corresponding
decrease in disk mass, resulting in a lower yield of lbGRBs compared to sbGRBs.}
\blue{
Ultimately, future observations of binary neutron stars through
gravitational waves and radio surveys will be crucial in refining the
neutron star mass function for both the primary and secondary
components (if they differ). This will provide a more robust
foundation for our analysis.}

\bigskip 

\noindent
\blue{\em (e) Dependence on the observational data}\\
\blue{
If the accumulation of future data were to alter the current lbGRB/sbGRB ratio, it would undoubtedly impact our inferences on \( M_{\rm ls} \). The direction of this effect is evident in Fig.~\ref{fig:GRBs} (as well as in all the figures in the appendix). An increase in the relative number of long cbGRBs would weaken our conclusions, whereas a rise in the relative number of short cbGRBs would strengthen them.
  }

\bibliography{biblio}

\end{document}